\documentclass[twocolumn]{aastex631}
\usepackage{float}
\usepackage{import}
\usepackage{color}
\usepackage{listings}
\usepackage{amsmath}
\usepackage{graphicx}
\usepackage[normalem]{ulem}
\usepackage{gensymb}

\graphicspath{{./}{figures/}}

\begin{document}

\shortauthors{Molina, Blunt \& Wang}
\title{Jointly Modeling Roman Coronagraph Astrometry and Photometry Improves Orbital Parameter Estimates}

\author[0009-0008-2430-6035]
{Farrah Molina}
\affiliation{Center for Interdisciplinary Exploration and Research in Astrophysics (CIERA), Northwestern University, Evanston, IL 60208, USA}
\affiliation{Department of Physics and Engineering, University of St. Thomas, Houston, TX, USA}

\author[0000-0002-3199-2888]{Sarah Blunt}
\affiliation{Department of Astronomy and Astrophysics, University of California, Santa Cruz, CA, USA}
\affiliation{Center for Interdisciplinary Exploration and Research in Astrophysics (CIERA), Northwestern University, Evanston, IL 60208, USA}

\author[0000-0003-0774-6502]
{Jason J. Wang}
\affiliation{Center for Interdisciplinary Exploration and Research in Astrophysics (CIERA), Northwestern University, Evanston, IL 60208, USA}

\begin{abstract}
Launching in 2027, the Nancy Roman Grace Space Telescope (Roman) has the potential to directly image exoplanets in reflected light for the first time. Roman imaging will introduce new constraints on exoplanet orbital parameters, since reflected-light intensity depends on orbital phase. In this Note, we discuss an addition to the open-source Python package \texttt{orbitize!}, which allows users to model exoplanet orbits using joint constraints from astrometry and photometric variations due to orbital phase. To investigate the impact of adding photometric data into the orbital model, we simulated realistic measurements of partial orbits, both including and excluding photometry in our model, and computed orbital posteriors. We found that fitting both astrometry and photometry improves posterior precision relative to fitting astrometry alone. This effect was more pronounced for higher-SNR images; for example, photometric data with SNR=10 yielded 33\%  improvement in inclination precision when including photometry, while SNR=3 data yielded only 12\% improvement. 

\end{abstract}

\section{Introduction}
\label{sec:intro}

The Habitable Worlds Observatory (HWO) will aim to directly image exoplanets in nearby star systems to spectroscopically characterize Earth-mass planets in the habitable zones of Sun-like stars \citep{Decadal}. To do this, the imaging community needs to build an instrument capable of achieving a contrast of 1e-10 in visible reflected light. 
All directly-imaged exoplanets have been detected in infrared, with current state-of-the-art space-based instruments achieving an intensity contrast of $\sim$1e-5-1e-6 \citep{JWST}. The Roman Coronagraph Instrument (CGI) is a technology demonstration intended to bridge this gap (\citealt{CGI}). After launch, time has been allocated for CGI to attempt to directly image an exoplanet in visible reflected light for the first time. The Roman CGI is expected to achieve an intensity contrast of $\sim$1e-8-1e-9 \citep{CGI} as a technological demonstration for the HWO mission. Repeated CGI imaging will allow us to see the planet's reflected-light intensity change as it orbits its host star. This is distinct from current direct imaging, which detects emitted light, and therefore is not sensitive to phase variations.
In theory, observing this change in intensity will allow us to place additional constraints on a planet's orbit using its photometric variability. In this Note, we explore how jointly modeling CGI photometry and astrometry may affect our knowledge of a planet's orbit, and how the signal noise ratio (SNR) affects the precision of the recovered orbital parameter posterior.

To investigate the impact of phase-dependent photometry on orbital parameter recovery, we modified \texttt{orbitize!}, an open-source software toolkit for Bayesian orbit modeling of directly-imaged stellar companions (\citealt{orbitize!}, \citealt{orb2024}). The structure of this Note is as follows: Section 2 outlines a suite of simulations we ran using \texttt{orbitize!}, and Section 3 analyzes the results of the simulations and discusses future work. 

\section{Simulations} 
\label{sec:simu}

\subsection{Mock Data}
  In order to investigate the impact of SNR on recovered orbital posteriors, we generated three mock data sets of astrometry and photometry. Following the definitions in \cite{orbitize!}, our test parameters are as follows: semimajor axis = 50 AU, eccentricity = 0.3, inclination = 30\textdegree, $\Omega$ = 60\textdegree, $\omega$= 120\textdegree, stellar parallax = 30 mas, albedo = 0.5, stellar mass = 1.25 M$_{\odot}$, and $\tau_{58849}$ = 0.

To test the impact of jointly modeling astrometry and reflected-light photometry, we modified \texttt{orbitize!}, a Bayesian tool for computing the posteriors over orbital elements of directly imaged planets \citep{2020AJ....159...89B}. Our implementation assumes a Lambertian disk reflection model (see Appendix), which describes ``matte'' reflection, or equally distributed reflection, off of a rough, porous surface. The assumption of a Lambertian disk is fixed in the current version of \texttt{orbitize!}; future work will focus on creating a model that can accommodate more detailed reflection laws. We used \texttt{orbitize!} to predict astrometric and photometric values, then assigned uncertainties assuming photon noise limited observations for SNRs of 3, 5, and 10. We assumed $\lambda=500\:$nm and D$=2.4\:$m, the diameter of Roman's CGI mirror. In other words, we defined the astrometric and photometric uncertainty as: 

\begin{equation}
   \mathrm{astrometric \;uncertainty} =\frac{\lambda/\rm D}{\rm SNR} 
\end{equation} 

\begin{equation}
   \mathrm{photometric \;uncertainty} =\frac{\rm photometric \;value}{\rm SNR}. 
\end{equation} 

We also randomly added Gaussian noise to each measurement. 

\subsection{Posterior Recovery}
Having generated our mock data sets, we next used \texttt{ptemcee} \citep{vousden, foreman}, a parallel-tempered Affine-invariant Markov Chain Monte Carlo (MCMC) sampling algorithm, to recover posterior distributions. We used \texttt{orbitize!} version 3.3.0 to perform these tests. Each test used 20 temperatures, 1000 walkers, 10,000 burn-in steps, and 100,000 total steps per walker. We thinned each resulting posterior by saving 1 out of every 10 steps. We ran two orbit fits for each SNR, one test using only the astrometry data, the other using both astrometry and photometry data. We assessed MCMC convergence by eye. 


\section{Results}
\label{sec:results}
\begin{figure*}
    \centerline{\includegraphics[width=0.75\linewidth]{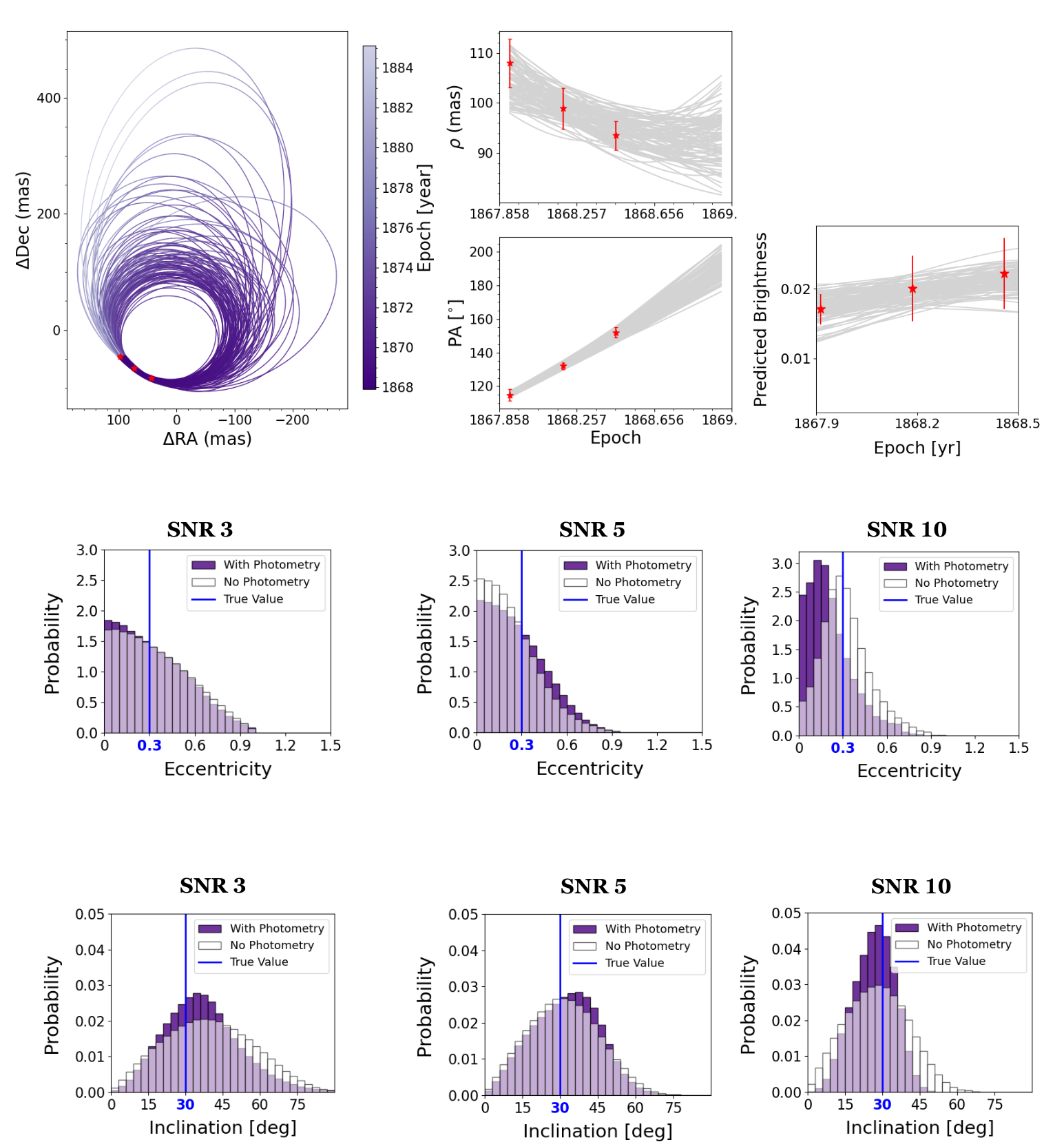}}
    \caption{The top four panels show 100 random draws from the orbital posterior for our SNR=10 case, jointly fitting astrometry and photometry. The bottom six panels show the posterior orbital constraints with and without photometric measurements for three different SNRs. The true values for the eccentricity and inclination are bolded and outlined in blue to show the improvements in constraint with an increasing SNR value. Takeaway: including photometry generally improves eccentricity and inclination constraints, and the improvement is more pronounced for higher values of SNR.
    }
    \label{fig:myfigure1}
\end{figure*}

Figure \ref{fig:myfigure1} visualizes the results of one joint photometry+astrometry fit, assuming SNR=10. In this case, including photometry improves the posterior constraint on inclination by 33\%. The bottom panels of this figure show the impact of SNR on posterior constraints. As expected, increasing SNR generally improves posterior constraints on both eccentricity and inclination, both when fitting only astrometry and astrometry+photometry. However, the value added from photometry increased with SNR; for SNR=3, including photometry improved the inclination posterior constraint by 12\%, while for SNR=10, including photometry improved the inclination posterior constraint by 33\%. 

\subsection{Conclusion}
\label{sec:subsection2label}

This project demonstrates the quantifiable impact of including photometric measurements with reasonable uncertainties expected for Roman CGI in orbit fits. We observe that these improvements are dependent on SNR, with a higher SNR producing a greater improvement compared to the astrometric-only posteriors. These results will allow for more accurate predictions of orbital parameters of directly imaged planets, in turn contributing to the characterization of exoplanet populations.

Future work for this project could focus on replacing the simplistic Lambertain disk reflection model with more realistic atmosphere models, investigating how posterior improvement depends on the true underlying orbital parameters, and investigating the relative impact of photometric and precursor RV or absolute astrometry constraints (e.g. from Gaia or Hipparcos) on the recovered orbits.  

\section*{Acknowledgements}

We thank the BOBA group at Northwestern University for their feedback in various stages of research. This material is based upon work supported by NASA under award 80NSSC24K0087.

\appendix
\label{sec:appendix}
For convenience, below we develop the equations for a Lambertain disk model in the \texttt{orbitize!} coordinate system. The absolute distance between a planet and star at true anomaly $f$ is given by:

\begin{equation}
    R =\frac{a(1-e^2)}{(1+e\cos{f})},
\end{equation}

and the line-of-sight distance, $z$, between the star and planet is:

\begin{equation}
    z = R(-\cos{\omega}\sin{i}\sin{f}-\cos{f}\sin{i}\sin{\omega}).
\end{equation}

We can use these two physical distances to obtain $\alpha$, the angle subtended by the path between the star, the planet, and the observer:

\begin{equation}
    \beta = \arctan\left(\frac{-R}{z}\right) + \pi
\end{equation}

\begin{equation}
    \alpha = \frac{1}{\pi}
    \left[\sin(\beta)+(\pi-\beta)\cos(\beta)\right].
\end{equation}

Finally, we obtain the equation that calculates the relative brightness of the planet over time:
\begin{equation}
    \mathrm{planet\ brightness} \propto \frac{\rm A \alpha}{\rm R^2}
\end{equation} where A is the planetary albedo, and R is the planetary radius. In the current implementation of \texttt{orbitize!}, we fit only for relative changes in brightness.

\software{\texttt{orbitize!} v3.4.0 (\citealt{orbitize!}, \citealt{orb2024}), \texttt{astropy} (\citealt{astropy:2013}, \citealt{astropy:2018}, \citealt{astropy:2022}) \texttt{numpy} \citep{numpy}, \texttt{matplotlib} \citep{matplotlib}, \texttt{emcee}  \citep{collette_python_hdf5_2014, emcee-Foreman-Mackey-2013, emcee_10996751}, \texttt{pandas}  \citep{pandas}, \texttt{HDF5}  \citep{h5py_7560547}.}

\bibliographystyle{aasjournal}
\bibliography{main}

@BOOK{Decadal,
       author = {{National Academies of Sciences, Engineering, and Medicine}},
        title = "{Pathways to Discovery in Astronomy and Astrophysics for the 2020s}",
         year = 2021,
          doi = {10.17226/26141},
       adsurl = {https://ui.adsabs.harvard.edu/abs/2021pdaa.book.....N}
}

@article{ JWST,
	author = {{Boccaletti, A.} and {Cossou, C.} and {Baudoz, P.} and {Lagage, P. O.} and {Dicken, D.} and {Glasse, A.} and {Hines, D. C.} and {Aguilar, J.} and {Detre, O.} and {Nickson, B.} and {Noriega-Crespo, A.} and {Gáspár, A.} and {Labiano, A.} and {Stark, C.} and {Rouan, D.} and {Reess, J. M.} and {Wright, G. S.} and {Rieke, G.} and {Garcia Marin, M.} and {Lajoie, C.} and {Girard, J.} and {Perrin, M.} and {Soummer, R.} and {Pueyo, L.}},
	title = {JWST/MIRI coronagraphic performances as measured on-sky},
	DOI= "10.1051/0004-6361/202244578",
	url= "https://doi.org/10.1051/0004-6361/202244578",
	journal = {A\&A},
	year = 2022,
	volume = 667,
	pages = "A165",
}

@ARTICLE{orb2024,
       author = {{Blunt}, Sarah and {Wang}, Jason and {Hirsch}, Lea and {Tejada}, Roberto and {Nagpal}, Vighnesh and {Surti}, Tirth and {Covarrubias}, Sofia and {McKenna}, Thea and {Ch{\'a}vez}, Rodrigo and {Llop-Sayson}, Jorge and {Arora}, Mireya and {Chavez}, Amanda and {Cody}, Devin and {Choudhary}, Saanika and {Smith}, Adam and {Balmer}, William and {Stolker}, Tomas and {Gallamore}, Hannah and {{\'O}}, Clarissa and {Nielsen}, Eric and {De Rosa}, Robert},
        title = "{orbitize! v3: Orbit fitting for the High-contrast Imaging Community}",
      journal = {The Journal of Open Source Software},
         year = 2024,
        month = sep,
       volume = {9},
       number = {101},
          eid = {6756},
        pages = {6756},
          doi = {10.21105/joss.06756},
archivePrefix = {arXiv},
       eprint = {2409.11573},
 primaryClass = {astro-ph.IM},
       adsurl = {https://ui.adsabs.harvard.edu/abs/2024JOSS....9.6756B}
}

@ARTICLE{foreman,
       author = {{Foreman-Mackey}, Daniel and {Hogg}, David W. and {Lang}, Dustin and {Goodman}, Jonathan},
        title = "{emcee: The MCMC Hammer}",
      journal = {\pasp},
         year = 2013,
        month = mar,
       volume = {125},
       number = {925},
        pages = {306},
          doi = {10.1086/670067},
archivePrefix = {arXiv},
       eprint = {1202.3665},
 primaryClass = {astro-ph.IM},
       adsurl = {https://ui.adsabs.harvard.edu/abs/2013PASP..125..306F}
}

@ARTICLE{vousden,
       author = {{Vousden}, W.~D. and {Farr}, W.~M. and {Mandel}, I.},
        title = "{Dynamic temperature selection for parallel tempering in Markov chain Monte Carlo simulations}",
      journal = {\mnras},
         year = 2016,
        month = jan,
       volume = {455},
       number = {2},
        pages = {1919-1937},
          doi = {10.1093/mnras/stv2422},
archivePrefix = {arXiv},
       eprint = {1501.05823},
 primaryClass = {astro-ph.IM},
       adsurl = {https://ui.adsabs.harvard.edu/abs/2016MNRAS.455.1919V}
}

@ARTICLE{orbitize!,
       author = {{Blunt}, Sarah and {Wang}, Jason J. and {Angelo}, Isabel and {Ngo}, Henry and {Cody}, Devin and {De Rosa}, Robert J. and {Graham}, James R. and {Hirsch}, Lea and {Nagpal}, Vighnesh and {Nielsen}, Eric L. and {Pearce}, Logan and {Rice}, Malena and {Tejada}, Roberto},
        title = "{orbitize!: A Comprehensive Orbit-fitting Software Package for the High-contrast Imaging Community}",
      journal = {\aj},
         year = 2020,
        month = mar,
       volume = {159},
       number = {3},
          eid = {89},
        pages = {89},
          doi = {10.3847/1538-3881/ab6663},
archivePrefix = {arXiv},
       eprint = {1910.01756},
 primaryClass = {astro-ph.EP},
       adsurl = {https://ui.adsabs.harvard.edu/abs/2020AJ....159...89B}
}

@Article{matplotlib,
  Author    = {Hunter, J. D.},
  Title     = {Matplotlib: A 2D graphics environment},
  Journal   = {Computing in Science \& Engineering},
  Volume    = {9},
  Number    = {3},
  Pages     = {90--95},
  publisher = {IEEE COMPUTER SOC},
  doi       = {10.1109/MCSE.2007.55},
  year      = 2007
}

@INPROCEEDINGS{CGI,
       author = {{Kasdin}, N. Jeremy and {Bailey}, Vanessa P. and {Mennesson}, Bertrand and {Zellem}, Robert T. and {Ygouf}, Marie and {Rhodes}, Jason and {Luchik}, Thomas and {Zhao}, Feng and {Riggs}, A.~J. Eldorado and {Seo}, Byoung-Joon and {Krist}, John and {Kern}, Brian and {Tang}, Hong and {Nemati}, Bijan and {Groff}, Tyler D. and {Zimmerman}, Neil and {Macintosh}, Bruce and {Turnbull}, Margaret and {Debes}, John and {Douglas}, Ewan S. and {Lupu}, Roxana E.},
        title = "{The Nancy Grace Roman Space Telescope Coronagraph Instrument (CGI) technology demonstration}",
    booktitle = {Space Telescopes and Instrumentation 2020: Optical, Infrared, and Millimeter Wave},
         year = 2020,
       editor = {{Lystrup}, Makenzie and {Perrin}, Marshall D.},
       series = {Society of Photo-Optical Instrumentation Engineers (SPIE) Conference Series},
       volume = {11443},
        month = dec,
          eid = {114431U},
        pages = {114431U},
          doi = {10.1117/12.2562997},
archivePrefix = {arXiv},
       eprint = {2103.01980},
 primaryClass = {astro-ph.IM},
       adsurl = {https://ui.adsabs.harvard.edu/abs/2020SPIE11443E..1UK}

}

@ARTICLE{numpy,
       author = {{van der Walt}, St{\'e}fan and {Colbert}, S. Chris and {Varoquaux}, Ga{\"e}l},
        title = "{The NumPy Array: A Structure for Efficient Numerical Computation}",
      journal = {Computing in Science and Engineering},
         year = 2011,
        month = mar,
       volume = {13},
       number = {2},
        pages = {22-30},
          doi = {10.1109/MCSE.2011.37},
archivePrefix = {arXiv},
       eprint = {1102.1523},
 primaryClass = {cs.MS},
       adsurl = {https://ui.adsabs.harvard.edu/abs/2011CSE....13b..22V}
}

@article{astropy:2013,
  adsurl        = {http://adsabs.harvard.edu/abs/2013A\%26A...558A..33A},
  archiveprefix = {arXiv},
  author        = {{Astropy Collaboration} and {Robitaille}, T.~P. and {Tollerud}, E.~J. and {Greenfield}, P. and {Droettboom}, M. and {Bray}, E. and {Aldcroft}, T. and {Davis}, M. and {Ginsburg}, A. and {Price-Whelan}, A.~M. and {Kerzendorf}, W.~E. and {Conley}, A. and {Crighton}, N. and {Barbary}, K. and {Muna}, D. and {Ferguson}, H. and {Grollier}, F. and {Parikh}, M.~M. and {Nair}, P.~H. and {Unther}, H.~M. and {Deil}, C. and {Woillez}, J. and {Conseil}, S. and {Kramer}, R. and {Turner}, J.~E.~H. and {Singer}, L. and {Fox}, R. and {Weaver}, B.~A. and {Zabalza}, V. and {Edwards}, Z.~I. and {Azalee Bostroem}, K. and {Burke}, D.~J. and {Casey}, A.~R. and {Crawford}, S.~M. and {Dencheva}, N. and {Ely}, J. and {Jenness}, T. and {Labrie}, K. and {Lim}, P.~L. and {Pierfederici}, F. and {Pontzen}, A. and {Ptak}, A. and {Refsdal}, B. and {Servillat}, M. and {Streicher}, O.},
  doi           = {10.1051/0004-6361/201322068},
  eid           = {A33},
  eprint        = {1307.6212},
  journal       = {\aap},
  month         = oct,
  pages         = {A33},
  primaryclass  = {astro-ph.IM},
  title         = {{Astropy: A community Python package for astronomy}},
  volume        = 558,
  year          = 2013
}

@article{astropy:2018,
  author        = {{Astropy Collaboration} and {Price-Whelan}, A.~M. and {Sip{\H{o}}cz}, B.~M. and {G{\"u}nther}, H.~M. and {Lim}, P.~L. and {Crawford}, S.~M. and {Conseil}, S. and {Shupe}, D.~L. and {Craig}, M.~W. and {Dencheva}, N. and {Ginsburg}, A. and {Vand erPlas}, J.~T. and {Bradley}, L.~D. and {P{\'e}rez-Su{\'a}rez}, D. and {de Val-Borro}, M. and {Aldcroft}, T.~L. and {Cruz}, K.~L. and {Robitaille}, T.~P. and {Tollerud}, E.~J. and {Ardelean}, C. and {Babej}, T. and {Bach}, Y.~P. and {Bachetti}, M. and {Bakanov}, A.~V. and {Bamford}, S.~P. and {Barentsen}, G. and {Barmby}, P. and {Baumbach}, A. and {Berry}, K.~L. and {Biscani}, F. and {Boquien}, M. and {Bostroem}, K.~A. and {Bouma}, L.~G. and {Brammer}, G.~B. and {Bray}, E.~M. and {Breytenbach}, H. and {Buddelmeijer}, H. and {Burke}, D.~J. and {Calderone}, G. and {Cano Rodr{\'\i}guez}, J.~L. and {Cara}, M. and {Cardoso}, J.~V.~M. and {Cheedella}, S. and {Copin}, Y. and {Corrales}, L. and {Crichton}, D. and {D'Avella}, D. and {Deil}, C. and {Depagne}, {\'E}. and {Dietrich}, J.~P. and {Donath}, A. and {Droettboom}, M. and {Earl}, N. and {Erben}, T. and {Fabbro}, S. and {Ferreira}, L.~A. and {Finethy}, T. and {Fox}, R.~T. and {Garrison}, L.~H. and {Gibbons}, S.~L.~J. and {Goldstein}, D.~A. and {Gommers}, R. and {Greco}, J.~P. and {Greenfield}, P. and {Groener}, A.~M. and {Grollier}, F. and {Hagen}, A. and {Hirst}, P. and {Homeier}, D. and {Horton}, A.~J. and {Hosseinzadeh}, G. and {Hu}, L. and {Hunkeler}, J.~S. and {Ivezi{\'c}}, {\v{Z}}. and {Jain}, A. and {Jenness}, T. and {Kanarek}, G. and {Kendrew}, S. and {Kern}, N.~S. and {Kerzendorf}, W.~E. and {Khvalko}, A. and {King}, J. and {Kirkby}, D. and {Kulkarni}, A.~M. and {Kumar}, A. and {Lee}, A. and {Lenz}, D. and {Littlefair}, S.~P. and {Ma}, Z. and {Macleod}, D.~M. and {Mastropietro}, M. and {McCully}, C. and {Montagnac}, S. and {Morris}, B.~M. and {Mueller}, M. and {Mumford}, S.~J. and {Muna}, D. and {Murphy}, N.~A. and {Nelson}, S. and {Nguyen}, G.~H. and {Ninan}, J.~P. and {N{\"o}the}, M. and {Ogaz}, S. and {Oh}, S. and {Parejko}, J.~K. and {Parley}, N. and {Pascual}, S. and {Patil}, R. and {Patil}, A.~A. and {Plunkett}, A.~L. and {Prochaska}, J.~X. and {Rastogi}, T. and {Reddy Janga}, V. and {Sabater}, J. and {Sakurikar}, P. and {Seifert}, M. and {Sherbert}, L.~E. and {Sherwood-Taylor}, H. and {Shih}, A.~Y. and {Sick}, J. and {Silbiger}, M.~T. and {Singanamalla}, S. and {Singer}, L.~P. and {Sladen}, P.~H. and {Sooley}, K.~A. and {Sornarajah}, S. and {Streicher}, O. and {Teuben}, P. and {Thomas}, S.~W. and {Tremblay}, G.~R. and {Turner}, J.~E.~H. and {Terr{\'o}n}, V. and {van Kerkwijk}, M.~H. and {de la Vega}, A. and {Watkins}, L.~L. and {Weaver}, B.~A. and {Whitmore}, J.~B. and {Woillez}, J. and {Zabalza}, V. and {Astropy Contributors}},
  title         = "{The Astropy Project: Building an Open-science Project and Status of the v2.0 Core Package}",
  journal       = {\aj},
  year          = 2018,
  month         = sep,
  volume        = {156},
  number        = {3},
  eid           = {123},
  pages         = {123},
  doi           = {10.3847/1538-3881/aabc4f},
  archiveprefix = {arXiv},
  eprint        = {1801.02634},
  primaryclass  = {astro-ph.IM},
  adsurl        = {https://ui.adsabs.harvard.edu/abs/2018AJ....156..123A}
}

@software{pandas,
  author       = {The pandas development team},
  title        = {pandas-dev/pandas: Pandas},
  month        = jun,
  year         = 2026,
  publisher    = {Zenodo},
  version      = {v3.0.4},
  doi          = {10.5281/zenodo.21003741},
  url          = {https://doi.org/10.5281/zenodo.21003741},
}

\end{document}